# Full Daytime Sub-ambient Radiative Cooling with High Figure of Merit in Commercial-like Paints

Xiangyu Li,[1,2] Joseph Peoples,[1,2] Zhifeng Huang,[1,3] Zixuan Zhao,[1] Jun Qiu[1,4,5] and Xiulin Ruan[1,2,*]

[1] School of Mechanical Engineering, Purdue University, West Lafayette, IN, USA.
[2] Birck Nanotechnology Center, Purdue University, West Lafayette, IN, USA.
[3] School of Power and Mechanical Engineering, Wuhan University, Wuhan, China.
[4] School of Energy Science and Engineering, Harbin Institute of Technology, Harbin, China.
[5] State Key Laboratory for Digital Manufacturing Equipment and Technology, Huazhong University of Science and Technology, Wuhan, China.

*Correspondence: ruan@purdue.edu

**SUMMARY**
**Radiative cooling is a passive cooling technology by reflecting sunlight and emitting radiation in the atmospheric sky window. Although highly desired, full daytime sub-ambient radiative cooling in commercial-like single-layer particle-matrix paints is yet to be achieved. In this work, we have demonstrated full daytime sub-ambient radiative cooling in $CaCO_3$-acrylic paint by adopting large bandgap fillers, a high particle concentration and a broad size distribution. Our paint shows the highest solar reflectance of 95.5% among paints and a high sky-window emissivity of 0.94. Field tests show cooling power exceeding 37 W/m$^2$ and lower surface temperature more than 1.7˚C below ambient at noon. A figure of merit RC is proposed to compare the cooling performance under different weather conditions. The RC of our cooling paint is 0.62, among the best radiative cooling performance while offering unprecedented benefits of the convenient paint form, low cost, and the compatibility with commercial paint fabrication process.**

## Context & Scale

Radiative cooling is a passive cooling technology which not only can lower the cooling energy consumption but also holds great promise to mitigate the urban island effect and global warming. Recent studies report high performance by utilizing a reflective metal layer. However, metallic components can be prohibitive in a variety of commercial applications. In this work, we demonstrate full daytime sub-ambient radiative cooling in single-layer particle-matrix paints for the first time, with comparable or better performance than previous studies. This is enabled by several innovative approaches: $CaCO_3$ nanoparticle fillers to minimizes the absorption in the ultraviolet band; a high particle concentration with a broad particle size distribution to enable efficient broadband reflection of the sunlight. With high performance and great reliability, our paint offers a convenient, low-cost and effective radiative cooling solution.

## INTRODUCTION

Cooling represents a significant sector of energy consumption in both residential and commercial applications.[2] Passive radiative cooling can cool surfaces without any energy consumption, by directly emitting heat through a transparent spectral window of the atmosphere, from 8 µm to 13 µm (the "sky window"), to the deep sky which functions as an infinite heat sink with a temperature of 3K. If the thermal emission of the surface through the sky window exceeds its absorption of the sunlight, the surface can be cooled below the ambient temperature under direct sunlight. Compared to conventional air conditioners that consume electricity and only move heat from the inside of the space to the outdoors, passive radiative cooling not only saves power, but also combats global warming since the heat is directly lost to the deep space.

The use of passive radiative cooling can be dated back centuries,[3] and scientific studies on daytime radiative cooling began in the 1970s.[4] Paints with sub-ambient daytime radiative cooling capability have been pursued for long,[5,6,15,16,7-14] and TiO$_2$ particles, the common pigments in commercial paints, were used in most of these studies. The particle size was usually selected to be on the order of hundreds of nanometers to several microns, since light can be strongly scattered and reflected by particles with a size comparable to the wavelength.[17,18] A thin layer of TiO$_2$ white paint coated on aluminum demonstrated daytime below-ambient cooling during a winter day, and the high solar reflectance was attributed to the aluminum substrate.[5] Recently, full daytime sub-ambient cooling has been demonstrated in photonic structures and multilayers,[19,20] which renewed the interest in the development of various radiative cooling materials. A subset of the authors theoretically predicted full daytime sub-ambient cooling using a dual-layer structure with TiO$_2$-acrylic paint layer on top of a carbon black layer,[14] or utilizing a broad particle size distribution instead of a single size to enable efficient broadband scattering of sunlight.[21] However, these theoretical proposals have not been experimentally confirmed yet. In fact, the performance of TiO$_2$-acrylic paints are limited by solar absorption in the ultraviolet (UV) band due to the moderate 3.2 eV electron bandgap of TiO$_2$, and near-infrared (NIR) 0.7-3 µm band due to acrylic absorption[21]. Theoretical studies have indicated that the solar reflectance of TiO$_2$-acrylic paint is unlikely to exceed 92%.[14,21] Other materials were also explored, including SiC,[12] SiO$_2$[16] and ZnS.[22] Among these, a single-layer SiO$_2$ particle bed was developed and partial daytime sub-ambient cooling was demonstrated except for the noon hours.[16] A recent review paper on existing heat reflective paints summarizes that their solar reflectance is moderately high, around 80% to 90%, and none of the paints has achieved full daytime sub-ambient cooling.[13] On the other hand, porous polymers have recently been reported to accomplish daytime cooling[23,24] and they are paint-like, but there are several limitations compared to commercial particle-matrix paints, such as it is unclear if the pores can be maintained over a long period; the materials are considerably more expensive than commercial paints; and the thickness needs to be large. Other non-paint daytime radiative cooling solutions include polymer-metal dual layer,[25] silica nanocomposite-metal dual layer,[26] silica-metal dual layer,[27] and engineered wood.[28] Considering these existing studies, it is clear that it is still a pertinent and challenging task to create single layer commercial-like nanoparticle-matrix paint for daytime radiative cooling to gain wide commercial applications.

In this work we have experimentally demonstrated high solar reflectance, high sky-window emissivity, and full daytime sub-ambient radiative cooling in single-layer particle-matrix paints with strong performance. The work was included in a provisional patent filed on October 3, 2018 and a non-provisional international patent application (PCT/US2019/054566) filed on October 3, 2019 and published on April 9, 2020.[1] Our work is enabled by the use of several different approaches from commercial paints. To minimize the solar absorption in the UV band, we considered and investigated many alternative materials with higher electron band gaps, and achieved strong performance with CaCO$_3$-acrylic paint, where CaCO$_3$ has an electronic band gap >5 eV. To compensate its low refractive index[29] and enable strong scattering, we adopted a particle volume concentration of 60%, which is considerably higher than those in commercial paints. It is known for TiO$_2$ paints that the optical reflectance increases with the particle concentration up until around 30%, when the optical crowding effect often occurs and further increase of particle concentration would lead to overall decrease of the reflectance.[30] However, as the particle concentration continues to increase and pass the critical particle volume concentration (CPVC), the reflectance will increase again.[30,31] Hence, we selected the 60% concentration, which is well above CPVC. It also reduced the volume of acrylic and its absorption in the NIR band. Furthermore, a broad particle size distribution instead of a single size was used to efficiently scatter all wavelengths in the solar spectrum and hence enhance the solar reflectance, as predicted in our previous simulations.[21] Finally, the acrylic matrix introduces vibrational resonance peaks in the infrared band thus ensures a high sky window emissivity. Radiative properties characterizations and field tests of temperature drops and cooling powers have demonstrated strong performance of our paint.

## RESULTS AND DISCUSSION

CaCO$_3$-acrylic paint with 60% volume concentration was made using a commercially compatible process. A mixture of Dimethylformamide (DMF), acrylic and particle fillers was left fully dried. The thickness of the sample was around 400 µm to make the properties substrate-independent. The details are presented in the Experimental Procedures. The free-standing CaCO$_3$ nanoparticle-acrylic paint sample is shown in Figure 1A along with commercial white paint (DutchBoy Maxbond UltraWhite Exterior Acrylic Paint). The SEM image of the sample surface is shown in Figure 1B, where CaCO$_3$ fillers are rod-shape with length about 1.9 µm and diameter around 500 nm.

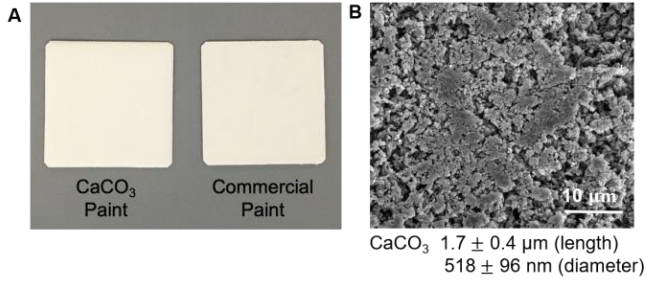

**Figure 1. CaCO₃-acrylic Paint**
(A) Our free-standing radiative cooling paint sample along with commercial white paint.
(B) An SEM image of the CaCO₃-acrylic paint. The particle size distribution is measured from SEM images.

The solar reflectance and sky window emissivity were then characterized. To achieve full daytime below-ambient cooling, it is critical to have both high solar reflectance and high sky-window emissivity. The solar reflectance is mainly contributed by fillers, and the sky-window emission can come from both the matrix and fillers, as shown in Figure 2A. High solar reflectance results from a combination of several factors including filler refractive index, volume concentration, particle size and size distribution. In the sky window, phonon (from the filler) or vibrational (from the polymer matrix) resonance peaks are essential for passive radiative cooling. At those peaks, photons are absorbed while interacting with phonons or vibrons, leading to high absorptivity and emissivity. The emissivity from 250 nm to 20 μm is shown in Figure 2B. While maintaining a similarly high emissivity 0.94 in the sky window with the commercial paint, our CaCO₃-acrylic paint reaches 95.5% reflectance in the solar spectrum due to lower absorption in the UV and NIR regions. This is much higher than the 87.2% reflectance of the commercial paint.

To help illustrate the physics behind the strong radiative properties, we performed photon Monte Carlo simulations of the nanoparticle composites with a thickness of 400 μm. Due to the ellipsoidal shape of the CaCO₃ particles, an effective particle size[32] was utilized with a diameter $\mu_d$ = 517.3 nm with $\sigma_d$ = 95.9 nm and length $\mu_l$ = 1744 nm with $\sigma_l$ = 408.4 nm. The uncertainty of the measurements was found to be ± 15.1 nm. The dielectric function of CaCO₃ was obtained from previous literature.[29] A modified Lorentz-Mie theory[21] was used to obtain the scattering coefficient $\eta_s$, absorption coefficient $\eta_a$, and asymmetric parameter of the nanoparticles in the matrix. A simple correction was then used to capture the dependent scattering effect due to the high concentration,[33]

$$\eta_{sd} = \eta_s(1 + 1.5f - 0.75f^2) \quad (1)$$
$$\eta_{ad} = \eta_a(1 + 1.5f - 0.75f^2) \quad (2)$$

where $f$ is the particle volume fraction, and $\eta_{sd}$ and $\eta_{ad}$ are the scattering and absorption coefficients after considering dependent scattering, respectively. These modified coefficients were then applied to a homogenous effective medium, where we used Monte Carlo method to solve the Radiative Transfer Equation by releasing 500,000 photons into the effective medium to predict reflectivity, absorptivity and emissivity. We covered 226 wavelengths from 250 nm to 2.5 μm. The results are shown in Figure 2C. We observed that using the single average particle size results in a solar reflectance of 90.9%, which is considerably lower than the experimental value. Including the size distribution increases the solar reflectance to 97.3% and agrees with the experimental value much better. The results support our previous theoretical proposal that multiple particle sizes can effectively scatter broadband wavelength.[21] The uncertainty from the Monte Carlo simulations is ± 0.3%. In addition, the effect of paint thickness is discussed in the Supplementary Note 1. Thinner films with thicknesses of 98, 131, and 177 μm still provide high solar reflectances of 88.9%, 93.4%, and 95.1%, respectively.

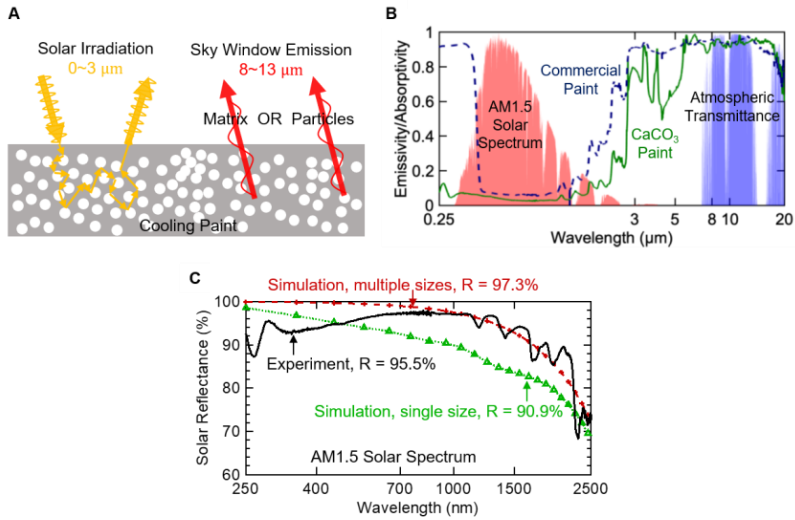

**Figure 2. Radiative Cooling Schematic, Spectral Characterization and Monte Carlo Simulations**
(A) For cooling paints exposed to direct solar irradiation, the fillers reflect sunlight between 0 to 3 µm, and the particles and/or polymer matrix emit in the sky window between 8 µm and 13 µm.
(B) The emissivity of our radiative cooling paint characterized from 0.25 µm to 20 µm compared with commercial white paint.
(C) Monte Carlo simulations on the solar reflectance of the $CaCO_3$-acrylic paint at 60% concentration compared with experimental results. Broad size distribution enhances the solar reflectance, and the results are in good agreement with the experimental measurement (black line).

A field test of surface temperature demonstrated full daytime cooling in West Lafayette, IN on March 21-23, 2018, as seen in Figure 3A. The sample stayed 10°C below the ambient temperature at night, and at least 1.7°C below the ambient temperature at a peak solar irradiation around 963 W/m$^2$. The relative humidity at 12 PM was around 40%. In another demonstration, a "P" pattern was painted with the $CaCO_3$-acrylic paint and the rest was painted with commercial white paint of the same thickness. It was then placed under direct sunlight. As shown in Figure 3B, the pattern was nearly invisible under a regular camera since both paints share similar reflectance in the visible spectrum (0.4-0.7 µm). However, it became much more distinctive under an infrared camera, clearly showing that the $CaCO_3$-acrylic paint was able to maintain a lower temperature under direct sunlight compared to the commercial white paint. The cooling power measurement using a feedback heater in Figure 3C showed an average cooling power of 56 W/m$^2$ during nights and 37 W/m$^2$ around noon (between 10 AM and 2 PM) in Reno, NV on August 1-2, 2018, when the relative humidity was 10% at 12 PM. A detailed analysis of the cooling power and comparison to experimental data are provided in the Supplementary Note 2.

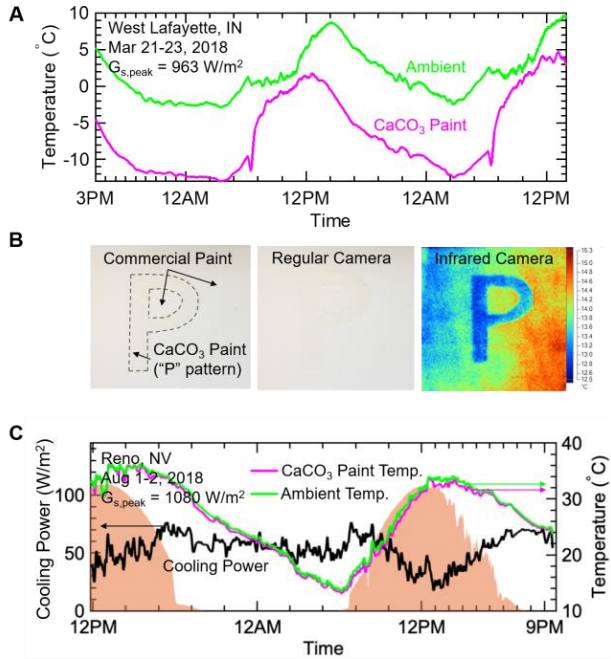

**Figure 3. Field Tests of the Cooling Paint**
(A) Field temperature measurement for the CaCO$_3$-acrylic paint over a period of more than one day.
(B) A "P" pattern painted with the CaCO$_3$-acrylic paint and the rest with the commercial white paint was placed under direct sunlight. The pattern was nearly invisible under a regular camera but became much more distinctive under an infrared camera due to the lower temperature of the CaCO$_3$-acrylic paint.
(C) Cooling power directly measured for the CaCO$_3$-acrylic paint using a feedback heater.
(A,C) The orange regions stand for the solar irradiation intensity.

In the literature[20,23,25-27] and thus far in this work, the cooling powers were reported for different locations and weather conditions, making it hard to fairly assess different radiative cooling systems. In fact, the weather conditions can critically affect the cooling power. For instance, humid weather can significantly reduce the cooling power compared to dry climates.[34] Here we define a simple figure of merit, RC, to help unify the radiative cooling performance by using the same ideal weather condition:

$$RC = \epsilon_{Sky} - r(1 - R_{Solar}) \qquad (3)$$

where $\epsilon_{Sky}$ is the emissivity in the sky window, $R_{Solar}$ is the total reflectance in the solar spectrum, and $r$ is the ratio of the solar irradiation power over the ideal sky window emissive power. The term $r(1 - R_{Solar})$ represents the ratio of the absorbed solar irradiation to the ideal sky window emissive power, and multiplying RC by the ideal sky window emissive power would yield the net cooling power. RC can be calculated to fairly evaluate different radiative cooling systems at the same solar irradiation and weather condition. We recommend to define a "standard figure of merit" using a standard peak solar irradiation of 1000 W/m$^2$ and surface temperature 300K, which would yield an ideal sky window emission power of 140 W/m$^2$ and a standard $r$ of 7.14. An ideal surface with 100% solar reflectance and an emissivity of 1 in the sky window has an RC of 1. The standard figures of merit for commercial white paint and our CaCO$_3$-acrylic paint come to be 0.02 and 0.62, compared to other state-of-the-art non-paint approaches as 0.41,[20] 0.64,[26] 0.44[27] and 0.68.[23] The RC of our CaCO$_3$-acrylic paint is comparable to or better than these reported radiative cooling systems. If the figure of merit is positive, the surface should be able to provide a net cooling effect. Unfortunately, the small RC of commercial white paint can be easily offset by unideal weather conditions, such as humidity, which is beyond the scope of the model. We note that a solar reflectance index (SRI) is widely used for paints.[35] Our RC is not intended to replace SRI, but to offer a complementary figure of merit specifically for assessing radiative cooling performance of different systems. With this definition, we suggest that there is no need to perform field tests for future material research on radiative cooling, since the ideal dry and summer days are only accessible for limited locations and time windows of the year. On the other hand, researchers who wish to explore the effects of weather conditions on radiative cooling can continue to perform field tests.

Abrasion tests were performed with Taber Abraser Research Model according to ASTM D4060.[36] A pair of abrasive wheels (CS-10) were placed on the surface with 250g load per wheel. The mass loss was measured every 250 cycles and the refacing was done every 500 cycles as required. The wear index ($I$) is defined as the weight loss in the unit of mg per 1000 cycles as

$$I = \Delta m \times 1000/C \quad (4)$$

where $\Delta m$ is the weight loss and $C$ is the cycle number. Using a linear fit to the mass loss, Figure 4A shows that the wear indexes of the commercial exterior paint and $CaCO_3$ paint are 104 and 84, respectively. Overall, our cooling paint showed similar abrasion resistance compared to commercial exterior paints. In a weathering test, the $CaCO_3$ paint was exposed to outdoor weathering including rain and snow for around 3 weeks. Figure 4B shows that the solar reflectance remained within the uncertainty level during the testing period. The sky window emissivity stayed 0.94 at the beginning and the end of the test. Moreover, the Supplementary Video of a running water test is provided to illustrate that our paint is water resistant. The viscosity of the $CaCO_3$ paint was measured and compared with water and oil-based commercial paints in Figure 4C. The viscosity of the commercial paints was retrieved from Whittingstall's work.[37] Our $CaCO_3$ paint showed lower viscosity, indicating that it can be brushed and dried similarly to commercial paints, as demonstrated by the Supplementary Video. The viscosity can be further adjusted by changing the type and amount of solvent used.

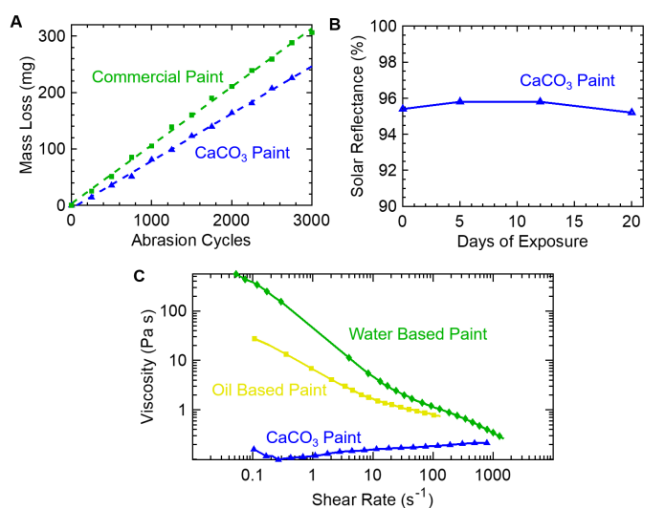

**Figure 4. Abrasion, Weathering and Viscosity Characterizations of the Cooling Paint**
(A) The mass loss as a function of the cycle number in abrasion tests. The slope retrieved was used to calculate the wear index.
(B) The solar reflectance remained the same during the 3-week outdoor weathering test.
(C) The viscosity of the cooling paint was compared to that of commercial paints. Commercial paint data is reproduced with permission.[37] 2011, TA Instruments Publication.

Due to the wide availability of $CaCO_3$ in natural minerals, the cost of our radiative cooling paints is anticipated to be comparable to or even lower than that of commercial white paints. The cost of the particle fillers for covering 100 m² area is less than $1.5 for $CaCO_3$ paint, making them among the most cost-effective radiative cooling solutions with full daytime below-ambient cooling capability. A detailed analysis of material costs and energy savings is given in the Supplementary Note 3.[38]

In this work, we have demonstrated that $CaCO_3$-acrylic paint, with high particle concentration and broad size distribution, is able to achieve full daytime below-ambient cooling with high figure of merit of 0.62. The intrinsic large band gap, appropriate particle size, broad particle size distribution and high particle concentration are all essential to strongly reflect the sunlight. The vibrational resonance peaks of the acrylic matrix provide strong emission in the sky window. Our radiative cooling paint showed cooling performance among the best of the reported state-of-the-art approaches, while offering unprecedented combined benefits, including convenient paint form, low cost, and the compatibility with commercial paint fabrication process. Future studies can aim at achieving higher performance with thinner films by exploring materials and structures. During the review process of our paper, we were made aware that results similar to part of our work, i.e., high solar reflectance in paints embedded with high concentration dielectric particles, were also reported in another paper.[39]

## EXPERIMENTAL PROCEDURES

### Sample Fabrication

The $CaCO_3$ particles were mixed with DMF first, followed by a 10-minute sonication to reduce particle agglomerations. We used Elvacite 2028 from Lucite International as the acrylic matrix for its low viscosity. The acrylic was then slowly added to the mixture to dissolve. The mixture was later degassed in vacuum chamber to remove air bubbles introduced during mixing and ultrasonication. The mixture was then poured in a mold and dried overnight till all solvent was gone. The dried paint was released from the mold as a free-standing layer with a thickness around 400 µm.

### Optical Measurement

The solar reflectance was measured on a Perkin Elmer Lambda 950 UV-VIS-NIR spectrometer with an integrating sphere, using a certified Spectralon diffuse reflectance standard. The uncertainty was $\pm 0.5\%$ based on the results of five different samples made in separate batches. Additional calibration was done with a silicon wafer, and the solar reflectance of the $CaCO_3$-acrylic paint was 94.9%, which is consistent with that from the diffuse reflectance standard. The IR measurements were done on a Nicolet iS50 FTIR spectrometer with a PIKE Technologies integrating sphere and the uncertainty of the PIKE Technologies diffuse reflectance standard was $\pm 0.02$.

### Field Test Setups

Two field test setups were made to characterize the cooling performance of the paints on the roof of a high-rise building, as shown in Figure 5. Figure 5A is a temperature test setup to demonstrate the below-ambient cooling capability. A cavity was made from a block of white styrofoam, providing excellent insulation from heat conduction. The 5 cm square sample was suspended above the styrofoam base. A thin layer of low-density polyethylene (LDPE) film was used as a shield against forced convection. T-type thermocouples were attached to the bottom of the samples for temperature measurement. Additional shaded thermocouple outside of the setup was used to monitor the ambient temperature. A pyranometer was placed to measure the solar irradiation, including both direct and diffuse components. According to the manufacturer, the accuracy depends on the angle of the irradiation, with directional errors less than 20 $W/m^2$ at 80 solar zenith angle. Figure 5B shows the cooling power test setup. A feedback heater was attached to the back of the sample to synchronize the sample and ambient temperatures. As the sample was heated to the ambient temperature, the power consumption of the heater was recorded as equal to the cooling power. The uncertainty, $\pm 5$ $W/m^2$, was calculated as the standard deviation of measured night-time cooling power at a stable surface temperature. The conduction and convection losses were insignificant (see details in Supplementary Note 2). Additionally, the setups were located on a wood board, which was further insulated on a metal table above the ground to avoid the heating effect from the ground (Figure 5C). Both setups were covered by silver mylar to reflect solar irradiation. The actual pictures of the two onsite cooling setups were shown in Figure 5D. The sidewalls and the bottoms of the styrofoam setups were enclosed by silver mylar to prevent additional radiation heat transfer.

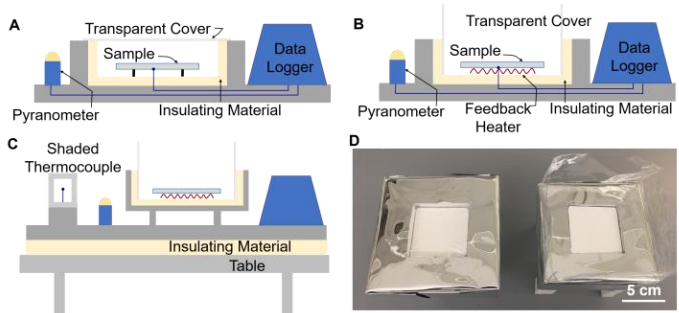

**Figure 5. Field Test Setups for Cooling Performance Characterization**
(A) A temperature test setup where the temperatures of the sample and the ambient were recorded. A lower sample temperature than the ambient indicated below-ambient cooling. (B) A cooling power test setup where the sample was heated to the same temperature as the ambient using a feedback heater. The power consumption of the heater was equal to the cooling power.
(C) The setups were further put on a high-rise table to eliminate the heating effect from the ground. A shaded thermocouple was located at the similar height with the cooling samples to avoid overheating from the ground.
(D) Pictures of actual setups. On the left and right are temperature and cooling power test setups, respectively.

**Monte Carlo Simulation**

The Monte Carlo simulations were performed in a similar manner as our previous work,[21] but a correction for dependent scattering was added. The photon packet is released at the top of the interface between air and the nanocomposite. The photon starts with a weight of unity and a normal direction to the air-composite interface. As the photon propagates across air-composite interface and through the medium it will lose weight scaling with the absorbing coefficient and will change direction affected by the scattering coefficient and asymmetric parameter. If the photon propagates to the bottom interface and makes it through, that weight is considered transmitted. If the photon propagates back up to the top surface and goes through the interface it is considered reflected.

## SUPPLEMENTAL INFORMATION
Supplemental Information includes three figures and a movie.

Figure S1. The Solar Reflectance of the $CaCO_3$-acrylic Paint with 60% Particle Concentration and Different Film Thicknesses, Supported by PET Films
Figure S2. The Ambient Temperature and the Sample Temperature Compared with the Sidewall Temperature of the Setup
Figure S3. The Theoretical Cooling Power Compared with Experimental Measurements of the $CaCO_3$ Paint
Movie S1. Drying and Water Running Test of the Cooling Paint

## AUTHOR CONTRIBUTIONS
Conceptualization, X.R.; Methodology, X.R. and X.L.; Software, J.P., Z.H. and J.Q.; Investigation, X.L., J.P. and Z.Z.; Writing – Original Draft, X.R. and X.L.; Writing – Review & Editing, X.R., X.L., J.P., Z.H, J.Q. and Z.Z.; Funding Acquisition, X.R.; Resources, X.R.; Supervision, X.R.

## ACKNOWLEDGMENTS
The authors thank Mian Wang, Jacob Faulkner, Professor George Chiu and Professor Zhi Zhou at Purdue University for their help on sample fabrication, and thank Professor Yan Wang at University of Nevada Reno for his assistance on the field tests. This research was supported by the Cooling Technologies Research Center at Purdue University and the Air Force Office of Scientific Research through the Defense University Research Instrumentation Program (Grant No. FA9550-17-1-0368).

## CONFLICTS OF INTEREST
X.R., X.L., Z.H. and J.P. are the inventors of a provisional patent application filed on October 3, 2018 and a non-provisional international patent application (PCT/US2019/054566) filed on October 3, 2019 and published on April 9, 2020,[1] which included the work described here.

## REFERENCES AND NOTES
1. Ruan, X., Li, X., Huang, Z., and Peoples, J.A. (2019). Metal-free Solar-reflective Infrared-emissive Paints and Methods of Producing the Same. PCT/US2019/054566, filed October 3, 2019. https://patentscope.wipo.int/search/en/detail.jsf?docId=WO2020072818&tab=PCTBIBLIO
2. Administration, U.S.E.I. (2018). Annual Energy Outlook 2018 with projections to 2050. J. Phys. A Math. Theor. *44*, 1–64.
3. Barker, R. (1857). Process of Making Ice in the East Indies. Sci. Am. *13*, 75–75.
4. Catalanotti, S., Cuomo, V., Piro, G., Ruggi, D., Silvestrini, V., and Troise, G. (1975). The radiative cooling of selective surfaces. Sol. Energy *17*, 83–89.
5. Harrison, A.W. (1978). Radiative cooling of TiO2 white paint. Sol. Energy *20*, 185–188.
6. Granqvist, C.G., and Hjortsberg, A. (1981). Radiative cooling to low temperatures: General considerations and application to selectively emitting SiO films. J. Appl. Phys. *52*, 4205–4220.
7. Andretta, A., Bartoli, B., Coluzzi, B., Cuomo, V., Andretta, A., Bartoli, B., Coluzzi, B., Selective, V.C., and For, S. (1981). Selective Surfaces for Natural Ccooling Devices. J. Phys. Colloq. *42*, C1-423-C1-430.
8. Lushiku, E.M., Hjortsberg, A., and Granqvist, C.G. (1982). Radiative cooling with selectively infrared-emitting ammonia gas. J. Appl. Phys. *53*, 5526–5530.
9. Lushiku, E.M., and Granqvist, C.-G. (1984). Radiative cooling with selectively infrared-emitting gases. Appl. Opt. *23*, 1835.
10. Orel, B., Gunde, M.K., and Krainer, A. (1993). Radiative cooling efficiency of white pigmented paints. Sol. Energy *50*, 477–482.
11. Diatezua, D.M., Thiry, P.A., Dereux, A., and Caudano, R. (1996). Silicon oxynitride multilayers as spectrally selective material for passive radiative cooling applications. Sol. Energy Mater. Sol. Cells *40*,

253–259.
12. Gentle, A.R., and Smith, G.B. (2010). Radiative heat pumping from the Earth using surface phonon resonant nanoparticles. Nano Lett. *10*, 373–379.
13. Pockett, J. (2010). Heat Reflecting Paints and a Review of their Advertising Material. CHEMECA 2010 Eng. Edge, 26-29 Sept., 1–13.
14. Huang, Z., and Ruan, X. (2017). Nanoparticle embedded double-layer coating for daytime radiative cooling. Int. J. Heat Mass Transf. *104*, 890–896.
15. Bao, H., Yan, C., Wang, B., Fang, X., Zhao, C.Y., and Ruan, X. (2017). Double-layer nanoparticle-based coatings for efficient terrestrial radiative cooling. Sol. Energy Mater. Sol. Cells *168*, 78–84.
16. Atiganyanun, S., Plumley, J., Han, S.J., Hsu, K., Cytrynbaum, J., Peng, T.L., Han, S.M., and Han, S.E. (2018). Effective Radiative Cooling by Paint-Format Microsphere-Based Photonic Random Media. ACS Photonics *5*, 1181–1187.
17. Mie, G. (1908). Beiträge zur Optik trüber Medien, speziell kolloidaler Metallösungen. Ann. Phys. *330*, 377–445.
18. Lorenz, L. (1890). Lysbevægelsen i og uden for en af plane Lysbølger belyst Kugle.
19. Rephaeli, E., Raman, A., and Fan, S. (2013). Ultrabroadband photonic structures to achieve high-performance daytime radiative cooling. Nano Lett. *13*, 1457–1461.
20. Raman, A.P., Anoma, M.A., Zhu, L., Rephaeli, E., and Fan, S. (2014). Passive radiative cooling below ambient air temperature under direct sunlight. Nature *515*, 540–4.
21. Peoples, J., Li, X., Lv, Y., Qiu, J., Huang, Z., and Ruan, X. (2019). A strategy of hierarchical particle sizes in nanoparticle composite for enhancing solar reflection. Int. J. Heat Mass Transf. *131*, 487–494.
22. Nilsson, T.M.J., and Niklasson, G.A. (1995). Radiative cooling during the day: simulations and experiments on pigmented polyethylene cover foils. Sol. Energy Mater. Sol. Cells *37*, 93–118.
23. Mandal, J., Fu, Y., Overvig, A., Jia, M., Sun, K., Shi, N., Zhou, H., Xiao, X., Yu, N., and Yang, Y. (2018). Hierarchically porous polymer coatings for highly efficient passive daytime radiative cooling. Science *362*, 315–319.
24. Leroy, A., Bhatia, B., Kelsall, C.C., Castillejo-Cuberos, A., Di Capua H., M., Zhao, L., Zhang, L., Guzman, A.M., and Wang, E.N. (2019). High-performance subambient radiative cooling enabled by optically selective and thermally insulating polyethylene aerogel. Sci. Adv. *5*, eaat9480.
25. Gentle, A.R., and Smith, G.B. (2015). A Subambient Open Roof Surface under the Mid-Summer Sun. Adv. Sci. *2*, 2–5.
26. Zhai, Y., Ma, Y., David, S.N., Zhao, D., Lou, R., Tan, G., Yang, R., and Yin, X. (2017). Scalable-manufactured randomized glass-polymer hybrid metamaterial for daytime radiative cooling. Science. *7899*, 1–9.
27. Kou, J., Jurado, Z., Chen, Z., Fan, S., and Minnich, A.J. (2017). Daytime radiative cooling using near-black infrared emitters. ACS Photonics *4*, 626–630.
28. Li, T., Zhai, Y., He, S., Gan, W., Wei, Z., Heidarinejad, M., Dalgo, D., Mi, R., Zhao, X., Song, J., et al. (2019). A radiative cooling structural material. Science *364*, 760–763.
29. Ghosh, G. (1999). Dispersion-equation coefficients for the refractive index and birefringence of calcite and quartz crystals. Opt. Commun. *163*, 95–102.
30. Diebold, M.P. (2014). Application of Light Scattering to Coatings: A User's Guide (Springer).
31. DuPont (2007). DuPont ™ Ti-Pure titanium dioxide - Titanium Dioxide For Coatings. 1–38. https://www.tipure.com/en-/media/files/tipure/legacy/titanium-dioxide-for-coatings.pdf.
32. Frisvad, J.R., Christensen, N.J., Jensen, H.W., Frisvad, J.R., Christensen, N.J., and Jensen, H.W. (2007). Computing the scattering properties of participating media using Lorenz-Mie theory. In ACM SIGGRAPH 2007 papers on - SIGGRAPH '07 (ACM Press), p. 60.
33. Kaviany, M. (2012). Principles of Heat Transfer in Porous Media (Springer New York).
34. Liu, C., Wu, Y., Wang, B., Zhao, C.Y., and Bao, H. (2019). Effect of atmospheric water vapor on radiative cooling performance of different surfaces. Sol. Energy *183*, 218–225.
35. ASTM E1980 - 11 Standard Practice for Calculating Solar Reflectance Index of Horizontal and Low Sloped Opaque Surfaces (2013).
36. D4060, Standard Test Method for Abrasion Resistance of Organic Coatings by the Taber Abraser (2010).
37. Whittingstall, P. (2011). Paint Evalution Using Rheology.
38. Akbari, H., Levinson, R., and Rainer, L. (2005). Monitoring the energy-use effects of cool roofs on California commercial buildings. Energy Build. *37*, 1007–1016.
39. Mandal, J., Yang, Y., Yu, N., and Raman, A.P. (2020). Paints as a Scalable and Effective Radiative Cooling Technology for Buildings. Joule *4*, 1350–1356.

# Supporting Information

**Full Daytime Sub-ambient Radiative Cooling Paint with High Figure of Merit in Commercial-like Paints**

*Xiangyu Li, Joseph Peoples, Zhifeng Huang, Zixuan Zhao, Jun Qiu and Xiulin Ruan\**

**Supplementary Note 1: Thickness-dependence of the solar reflectance**

Thinner $CaCO_3$-acrylic paint samples with 60% concentration were also fabricated. Since these films are more fragile as free-standing samples, they were deposited on polyethylene terephthalate (PET) film using a film applicator to control the wet film thickness. The transparent PET substrate has a negligible effect on the solar reflectance. The dry film thickness was measured with a coordinate measuring machine (Brown&Sharp MicroXcel PFX) and a digital indicator (Mitutoyo Absolute), by averaging various spots of the samples. The solar reflectance is shown in Figure S1. The paint still showed high solar reflectance of 95.1%, 93.4%, and 88.9% at a thickness of 177, 131, and 98 µm, respectively. Monte Carlo simulation results incorporating the same size distribution with the main text for different film thicknesses are represented by the dash lines in Figure S1, showing very good agreement with the experimental data. Future studies can aim at achieving higher performance with thinner samples.

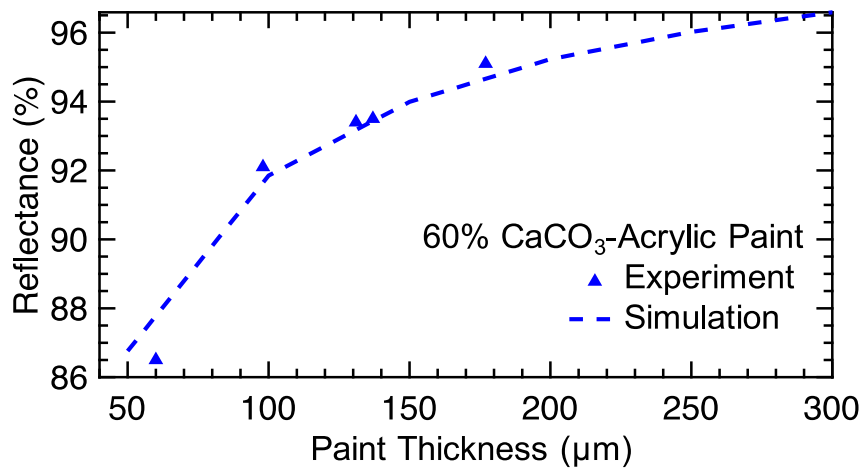

**Figure S1.** The solar reflectance of the $CaCO_3$-acrylic paint with 60% particle concentration and different film thicknesses, supported by PET films. The experimental results are represented by triangles, and the simulation results by the dash line. The transparent PET films have a negligible effect on the solar reflectance.

**Supplementary Note 2: Energy balance model for the cooling power characterization**

Before analyzing the results in the cooling power tests, we first justify that the conduction and convection losses were insignificant. To quantify the heat conduction loss, we recorded the temperature of the sidewall beneath the silver mylar enclosures, shown in Figure S2. The maximum temperature difference between the sidewall and the sample was less than 1.5°C, which translated to less than 5 $W/m^2$ of conduction loss, considering the thermal conductivity of styrofoam as 0.035 W/mK. During the noon hours, the silver mylar actually absorbed more sunlight than our sample, slightly reducing our cooling power during the day. Overall, the sample temperature was maintained the same as the ambient temperature with a feedback heater in direct cooling power measurements, hence the conduction and convection loss was minimized. We used T-type thermocouples with 0.127 mm diameter and 2 m long wires. Even if assuming a temperature difference as high as 2°C, the conduction loss through the thermocouple wires was estimated of only $10^{-6}$ $W/m^2$.

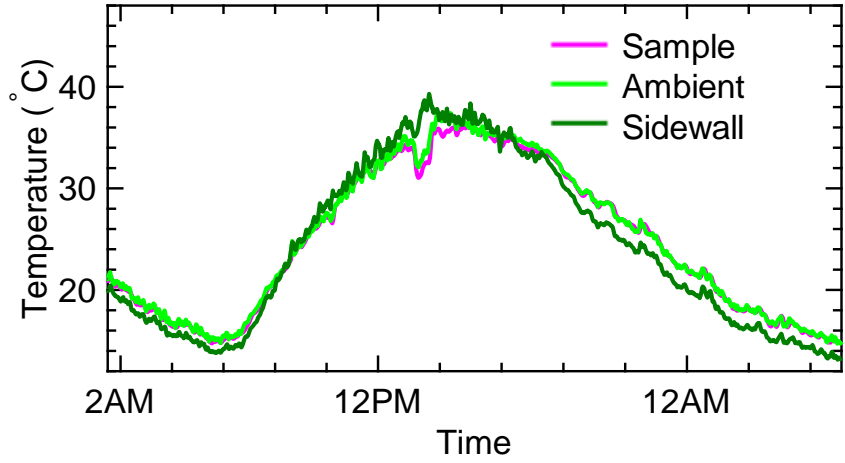

**Figure S2.** The ambient temperature and the sample temperature compared with the sidewall temperature of the setup. All three temperatures were close to each other, indicating a negligible conduction and convection contribution to the cooling power.

Using the temperature profile in the cooling power test, we analyzed the cooling power and compared to the experimental results for validation. The net cooling power $q''_{cooling}$ according to the energy balance model is

$$q''_{cooling} = \frac{mC_p}{A}\frac{dT}{dt} - \alpha G + q''_{radiation}(T) \tag{S1}$$

where $A$ is the surface area of the sample, $\frac{mC_p}{A}\frac{dT}{dt}$ accounts for the effect of heat capacity from the sample and the feedback heater, $\alpha$ stands for the solar absorption of the sample, $G$ represents the solar irradiation, $q''_{radiation}(T)$ is the radiation loss, and $q''_{cooling}$ is the net cooling power. $q''_{radiation}(T)$ term consists of the surface thermal emission through the sky window and radiation exchange with the ambient. The former term is a function of the sample surface temperature, and the latter term is neglected as the surface temperature is much closer to the ambient (~300K) than to the deep sky (~3K). Using the measured night-time cooling powers when $G = 0$, the radiation loss of the $CaCO_3$ paint was calibrated to be 60 W/m² at 15°C. Using the experimentally measured $\alpha$, $G$, $T$ and the calibrated radiation loss, the theoretical net cooling power was calculated and compared with experimental results of the heater power consumption in Figure S3. The agreement was reasonably well with a standard deviation around 10 W/m² without considering convection or conduction. This also indicates that the radiation loss to the deep sky and the power consumption of the feedback heater were the main heat fluxes in the system, confirming our assumption of negligible conduction and convection losses.

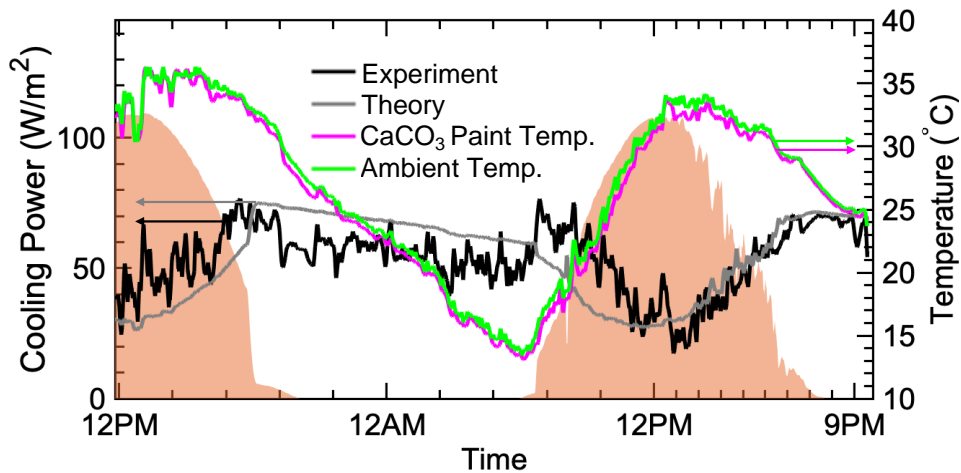

**Figure S3.** The theoretical cooling power compared with experimental measurements of the $CaCO_3$ paint. The radiation loss to the deep sky was calibrated to be 60 W/m² at 15°C. The model agrees reasonably well, confirming our assumption of negligible conduction and convection losses.

**Supplementary Note 3: Cost analysis of the cooling paint**

Due to the wide availability of $CaCO_3$ in natural minerals, the cost of our radiative cooling paint can be comparable to or even lower than that of commercial white paints. Calcite powders cost only 1 to 4 cents per kilogram, which is much cheaper than $TiO_2$ at around $ 1 per kilogram. The cost of the particle fillers for covering 100 m² area is less than $ 1.5 for $CaCO_3$ paint. With a similar fabrication method to the commercial paints, there is no additional cost for the manufacture process.

The Heat Island group at the Lawrence Berkeley National Lab has performed comprehensive tests with 85% solar reflectance roofing materials and concluded their energy savings to be 40 to 75 Wh/m$^2$/day on different buildings during one-month period in the summer.[37] Our cooling paints showed higher performance with over 95% solar reflectance, and with the solar irradiation around 5000 Wh/m$^2$/day, our paint can produce energy savings of 70 to 105 Wh/m$^2$/day, assuming the AC stock average efficiency to be 15. If the electricity cost is set to about $ 0.1 per kWh, the savings on the cooling cost will be $ 0.7/day for a moderate 100 m$^2$ apartment.